\documentclass{pasj00}
\draft

\begin{document}
\SetRunningHead{widget et al.}{WIDGET}
\Received{2010/07/05}
\Accepted{2010/10/21}

\title{WIDGET: System Performance and GRB Prompt Optical Observations}

%
 \author{
   Yuji \textsc{Urata}\altaffilmark{1},
   Makoto S. \textsc{Tashiro}\altaffilmark{2},
   Toru \textsc{Tamagawa}\altaffilmark{3,4},
   Fumihiko \textsc{Usui}\altaffilmark{5}, \\
   Makoto \textsc{Kuwahara}\altaffilmark{3,4},
   Hungmiao \textsc{Lin}\altaffilmark{1},
   Shoichi \textsc{Kageyama}\altaffilmark{3,4},
   Wataru \textsc{Iwakiri}\altaffilmark{2},\\
   Takako \textsc{Sugasahara}\altaffilmark{2},
   Kazuki \textsc{Takahara}\altaffilmark{2},
   Natsuki \textsc{Kodaka}\altaffilmark{2},\\
   Keiichi \textsc{Abe}\altaffilmark{2},
   Keisuke \textsc{Masuno}\altaffilmark{2},
   and    Kaori \textsc{Onda}\altaffilmark{2}
}

\altaffiltext{1}{Institute of Astronomy, National Central University, Chung-Li 32054, Taiwan}
\email{urata@astro.ncu.edu.tw}
\altaffiltext{2}{Department of Physics, Saitama University, Shimo-Okubo 255, Sakura, Saitama 338-8570}
\altaffiltext{3}{RIKEN (Institute of Physical and Chemical Research), 2-1 Hirosawa, Wako, Saitama 351-0198} 
\altaffiltext{4}{Tokyo University of Science, 1-3 Kagurazaka, Shinjyuku, Tokyo} 
\altaffiltext{5}{Japan Aerospace Exploration Agency, Institute of Space and Astronautical Science\\
  3-1-1 Yoshinodai, Chuo-ku, Sagamihara, Kanagawa 252-5210}


\maketitle

\begin{abstract}

  The WIDeField telescope for Gamma-ray burst Early Timing (WIDGET) is
  used for a fully automated, ultra-wide-field survey aimed at
  detecting the prompt optical emission associated with Gamma-ray
  Bursts (GRBs).  WIDGET surveys the {\it HETE-2} and {\it Swift}/BAT
  pointing directions covering a total field of view of
  $62^{\circ}\times62^{\circ}$ every 10 secounds using an unfiltered
  system.  This monitoring survey allows exploration of the optical
  emission before the $\gamma$-ray trigger. The unfiltered magnitude
  is well converted to the SDSS $r'$ system at a 0.1 mag level.
  Since 2004, WIDGET has made a total of ten simultaneous and one
  pre-trigger GRB observations.  The efficiency of synchronized
  observation with {\it HETE-2} is four times better than that of {\it
    Swift}. There has been no bright optical emission similar to that
  from GRB 080319B. The statistical analysis implies that GRB080319B
  is a rare event.
  This paper summarizes the design and operation of the WIDGET system
  and the simultaneous GRB observations obtained with this instrument.

\end{abstract}

\section{Introduction}

The prompt emission of Gamma-ray Bursts (GRBs) has become an
intriguing topic in modern astrophysics. In particular, the optical
flash associated with $\gamma$-ray emission poses significant
challenges both observational and theoretical.  In 1999, the robotic
instrument {\it ROTSE} successfully detected the first optical flash
from GRB990123 with a brightness that reached 8.9 mag 50 s after the
burst \citep{990123}.  Recently, the optical flash associated with
GRB080319B has been detected at high signal to noise using a number of
small telescopes \citep{080319b, raptor, bloom}, and could have been
observed without any instruments.  These are bright enough to be
detected by small aperture instruments.  Surprisingly, the bright
optical emission of GRB 080319B was also detected by the simple
weather monitoring equipment at Mt. Lemmon Observatory operated by
KASI in Korea (Urata et al., 2010 in prep).  In addition to these two
events, small robotic telescopes have also made prompt optical
observations during $\gamma$-ray activity such as of GRB041219A
\citep{041219a}, GRB050820A \citep{050820a}, GRB051111 \citep{051111}
and GRB061121 \citep{061121}.  Even though there were only six optical
detections before the $\gamma$-ray radiation died off, three major
emission models have been used to explain these optical emissions: (1)
reverse shock (e.g. \citet{re1,re2}): (2) synchrotron self Compton
(e.g. \citet{080319b}): and (3) simple synchrotron radiation from the
standard internal shock (e.g. \citet{shen}).  Furthermore,
\citet{yamazaki} predicts a possible optical precursor before the rise
of the $\gamma$-ray emission which would explain the plateau phase in
the X-ray afterglows.

The number of robotic telescopes responding to GRB detection alerts
through the GRB Coordinate Network (GCN) increased starting from the
{\it HETE-2} era.  These telescopes however are not sufficient to make
simultaneous and pre-trigger observations with prompt $\gamma$-ray
emission. Since they respond to the GRB position alert from
satellites, the delay time is always longer than several tens of
seconds.  Therefore, a $\gamma$-ray/X-ray precursor or longer duration
event is required to make optical observations simultaneous with
$\gamma$-rays.  In order to break through this limitation, we have
constructed and began operating the WIDeField telescope for Gamma-ray
Burst Early Timing (WIDGET) since 2005 \citep{tamagawa}. There are
three other similar instruments worldwide, the so-called ``Pi of the
Sky'' \citep{pi}, RAPTOR \citep{raptor} and TORTORA
\citep{tortora}. These three cover the sky in the Western hemisphere.
Only WIDGET covers the Eastern hemisphere.  In this paper, we report
on the development, performance and the first results of WIDGET.

\section{The {\it WIDGET} System}\label{system}

\subsection{Overview}
WIDGET is a fully robotic telescope system placed at the Kiso
astronomical observatory of the Institute of Astronomy, Faculty of
Science, University of Tokyo. The main aim of this system is to
monitor the {\it Swift} field-of-view and to detect GRB optical
flashes or possible optical precursors.  The prototype systems (called
WIDGET1 and WIDGET1.5) was operated at the Akeno observatory of the
Institute for Cosmic Ray Research University of Tokyo, which was used
for the Akeno Giant Air Shower Array (AGASA; Hayashida et al. 1996).
During the two year operation in Akeno, a fully robotic system was
established, and the system performance and site condition of Akeno
were evaluated (Tamagawa et al. 2005). System updates are summarized
in table \ref{system}.  Through these studies, we found that a dark
sky and a new improved optical system are required for more efficient
monitoring.  We had moved the system to the Kiso astronomical
observatory in 2006 November.  This was done to improve the sky
background level and weather conditions.  In addition, this allows us
to connect the WIDGET observations with monitoring of afterglows by
the 105cm Schmidt telescope at the same site without weather risk.
The telescope has been observing afterglows (e.g. \cite{020813,030329,
  040924,sum, 041006, 050319}).  In November 2006 we began the
installation of a new observatory hut and weather monitoring system
and performed testing. After three months of refining, full operation
started in February, 2007.

The current {\it WIDGET} system (hereafter WIDGET2) consists of four
optics systems, an automated observatory hut equipped with a cluster
of control computers and a house keeping system for unattended
service.  {\it WIDGET2} is powered by the commercial 100-volt
alternating current source and the commands and data are transmitted
via the digital service line (DSL).

\begin{figure}[htb]
  \begin{center}
    \FigureFile(160mm,80mm){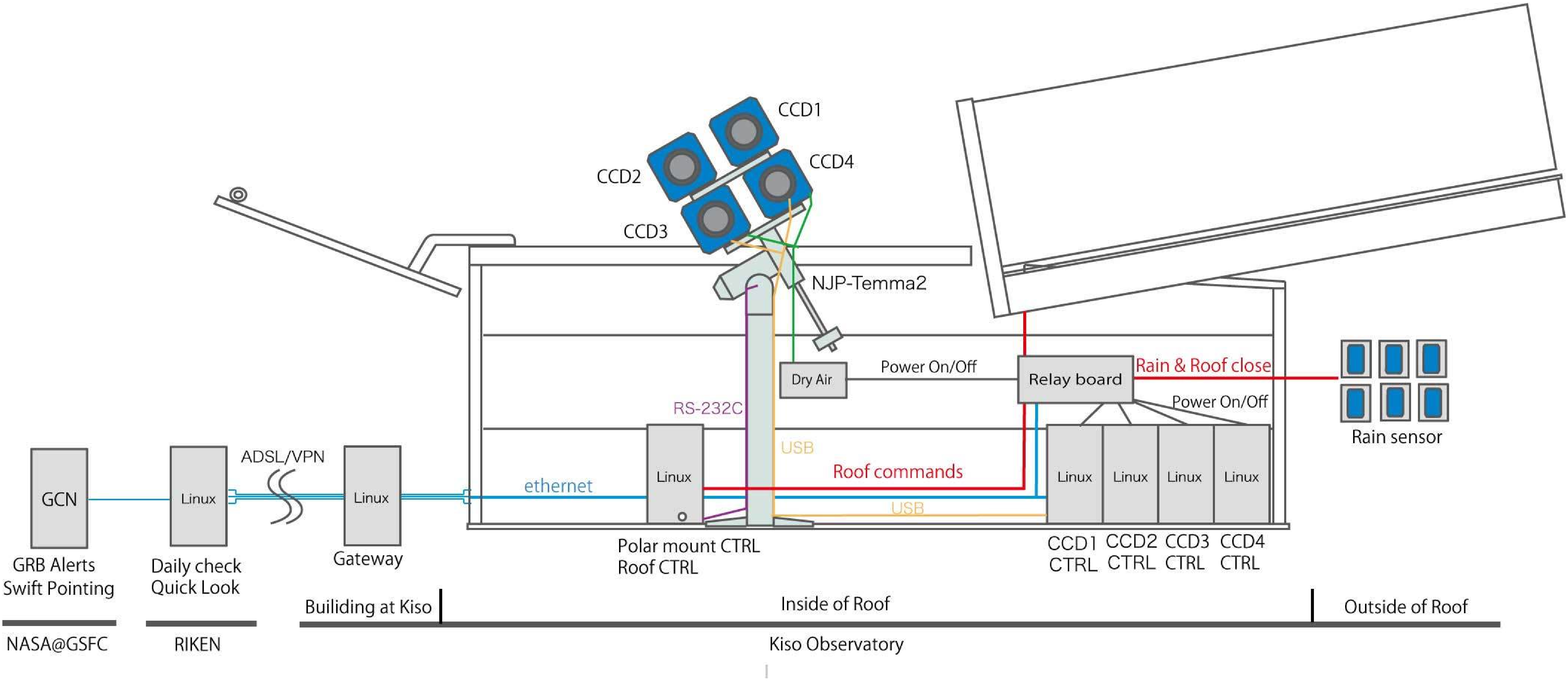}
  \end{center}
  \caption{Hardware block diagram of WIDGET.}\label{control}
\end{figure}

The telescope, with its related computers and electronics are housed
in the observatory hut with its one-direction sliding roof.  Figure
\ref{control} shows a schematic of the control system of cameras, the
monitoring system, the polar mount, and remaining equipment.  The
entire system is controlled by five computers and a relay board box
insde the observatory hut.

The time reference signal is provided by a network timing protocol
(NTP) server and a radio clock with an average error of less than 1 s.
The power supply for equipment such as CCD cameras, polar mounts,
control PCs can be turned off/on remotely via relay boards.

\subsection{Observational Hut}

\begin{figure}
  \begin{center}
    \FigureFile(160mm,80mm){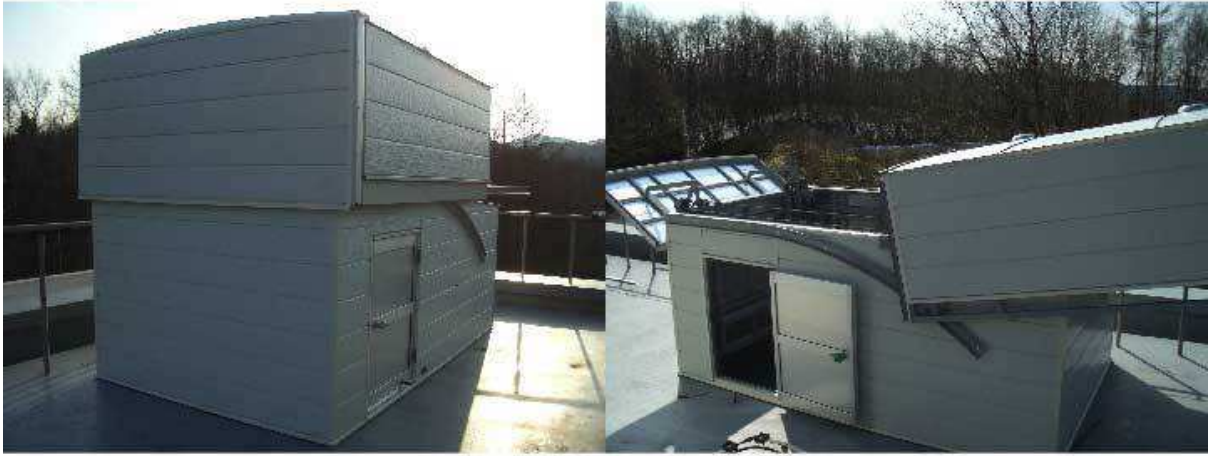}
  \end{center}
  \caption{The observing huts of WIDGET.}\label{fig:roof}
\end{figure}

The observatory hut is custom built by Human Comm Co. Ltd.. Its
dimensions are H $\times$ W $\times$ L $= 2 \times 2 \times 3$ meters.
As shown in Figure \ref{fig:roof}, the sliding roof on the optics side
the sliding roof dives down when in observation mode so as not to
interfere with the field of view of optics down to an altitude of
20$^\circ$.
The power line is connected and controlled by the relay boards, which
accept commands from a control PC via a socket connection.  The roof
is nominally opened and closed by preplanned commands, although if the
weather station detects rainfall the roof can be closed by the
automatically generated command.
Two small air ventilator fans made for sailboat usage are
installed on top of the roof. The power is supplied by the system's own solar
battery.  The temperature beneath the roof remains
nominal even on sunny summer days.

\subsection{Weather stations}

The weather sensing equipment placed outside and inside the hut
monitors rainfall, air temperature, wind speed and humidity. The
opening and closing of the sliding roof at night is fully dependent on
rain detection by the rain sensor.  There are a total of eight rain
sensor sets mounted around the roof to enable detection of rain
regardless of the direction it come from.  The inclination of the rain
sensors could have some dependence on the direction. For the
preservation of the instruments inside the observing hut, the roof
closes immediately when any of the sensors detects rain fall. To avoid
the chattering that occurs under misty conditions or light rainfall
that could lead to frequent opening and closing of the roof, we set
the hardware time delay to 30 minutes after the changing of the rain
status (rain to fine).

\subsection{Optics}

\begin{figure}
  \begin{center}
    \FigureFile(80mm,80mm){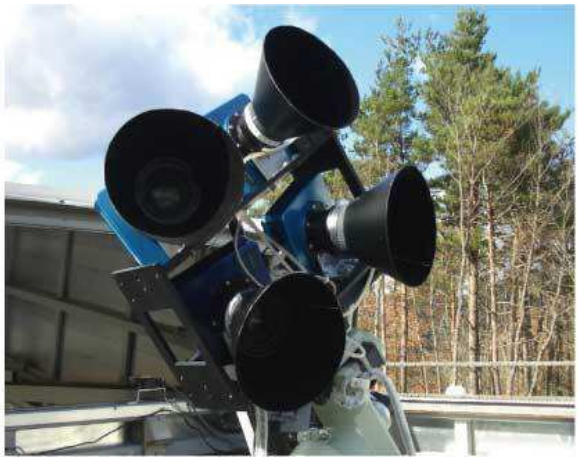}
  \end{center}
  \caption{The WIDGET optics attached to a polar mount.}\label{optics}
\end{figure}

\begin{figure}
  \begin{center}
    \FigureFile(160mm,160mm){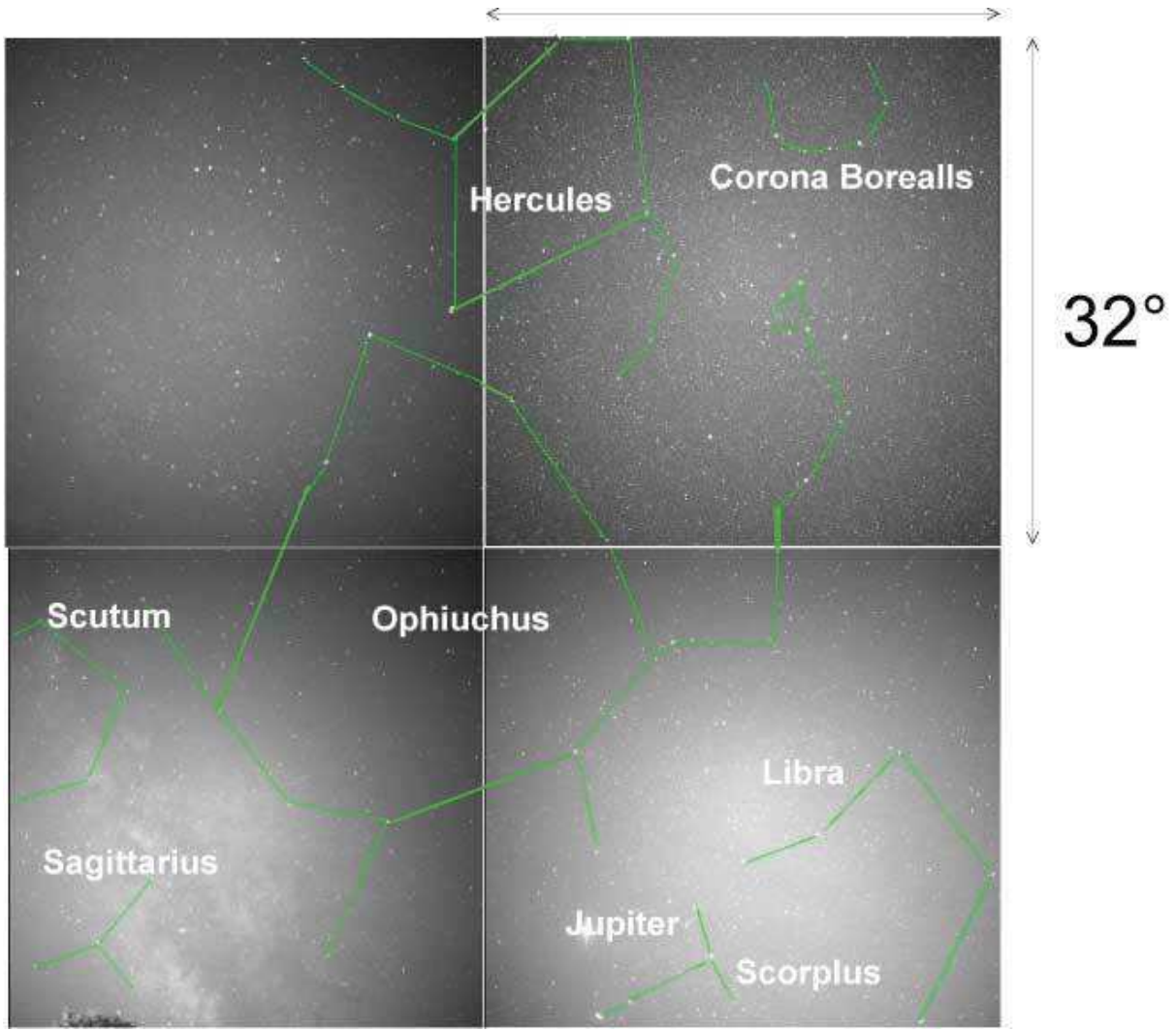}
  \end{center}
  \caption{The entire field of view of the WIDGET-2.}\label{widget-fov}
\end{figure}

The optics system consists of four sets of commercial CCD camera and
lens sets.  We employ Apogee Alta U10 CCD camera carrying a 2048
$\times$ 2048 format Atmel THX7899 CCD chip of 14-micron pixels. The
CCD chip is the front-illuminated type with a quantum efficiency of 38
\% for 720 nm optical light.  As shown in figure \ref{optics}, the
camera body is compact ($15\times15\times6.35$ cm) and light (1.4 kg)
enough to allow the installation of an array of four cameras on a
single small polar mount.  Each camera is equipped with a 50 mm f/1.2
Canon EF lens.  Each lens has a special lens baffle to block other
light sources from the ground and to prevent water condensations
during observing runs. A continuous flow of dry air is also provided
to the surface of the lens to avoid water condensations.  This
configuration yields a 32$^\circ \times$ 32$^\circ$ field of view for
each segment. As shown in Figure \ref{widget-fov}, the total field of
view with 2$^\circ$ field overlap between each segment is 62$^\circ
\times$ 62$^\circ$ which covers about 33\% of the {\it Swift}/BAT
observing field.  The electronics on-board the CCD camera have been
optimized to provide fast readout of the entire array within 5 s with
a 16 bit AD converter through an USB2.0 line.  The standard exposure
during operation is 5 s, so that we obtain sky images every 10 s
including the readout time.  This exposure cycle provides an optimal
balance between sensitivity and time-resolution.

There is a thermoelectric cooler with forced air control for the CCD
cooling.  Since the cooler delivers a maximum temperature difference
of $45^\circ$C below the ambient temperature, the actual temperature
depends on the season. The minimum and maximum are $-40^\circ$C in
winter and $-15^\circ$C in summer.  Bias and dark analysis confirms
the read out noise and dark level are 14.7e$^{-}$ and 1.1
e$^-$~pixel$^{-1}$~s$^{-1}$ at $-15^{\circ}$C, respectively. Hence,
the read out noise is about three times larger than the dark level
with the daily standard exposure time.

\subsection{Polar-mount}

As shown in figure 2, the optics is attached to a Takahashi NJP
Temma-2 polar mount. The polar mount is controlled via a serial line
by a PC running Linux OS with custom control software.  The maximum
slewing speed is 90 arcmin~s$^{-1}$, which is sufficient to make
effective monitoring observations for the {\it Swift's} monitoring
fields.  Due to the shape of the mount with the camera array, there
are some areas where the optics hit the pillar of the mount. Manual
inspection shows just where these areas are. Although these areas do
not affect normal operation, with imposed software operational limits.

\section{Daily operation and observations}

The daily operation is fully managed by automated observing
scripts. Each script controls the observation huts, mount, cameras and
weather monitoring system individually. To store observational data in
order, the CCD operation script gathers pointing information and
coordinates, and records them in the fits header unit.  Daily
operating status, including weather station status can be checked via
the internal web site which is updated every 5 minutes. The stored
data is summarized and transmitted via the Internet with house keeping
data and operation reports every morning.  Email notification is sent
to a human operator in case of problems related to the observation
huts.  The duty scientist checks the system and observation status
once a day.  The telescope automatically observes the sky every night
that is not rainy.  Thirty dark frames are obtained before and after
monitoring observations. Such automatic observations offer maximum
monitoring efficiency for observations prior to and simultaneous with
the GRB observation even under partially cloudy condition. The
observation field is selected based on the ``{\it Swift} Pointing
Dir'' alert from the GRB Coordinate Network (GCN). When the {\it
  Swift} pointing direction is not observable, such as for southern
sky targets (Dec$>-30^\circ$) or due to objects obstructing the
horizon at the Kiso observatory, WIDGET observes the zenith direction
without sidereal tracking. These images are stacked using median
filtering to obtain sky flat frames.  The previous systems (WIDGET1.0
and 1.5) observed mainly the {\it HETE-2} field of view and the
zenith.  WIDGET2 is monitoring these fields by repeated unfiltered 5 s
exposures.

The WIDGET2 system has no-quick response mode like the robotic
follow-up system, such as {\it ROTSE} or {\it RAPTOR}.  Once a GRB
occurs, WIDGET observes the field, until receiving the next {\it
  Swift} pointing information or until morning.  When {\it Swift}
detects a new GRB, the duty scientist is notified by cellphone. The
duty scientist checks the weather condition at the Kiso Observatory as
well as the overlap between the WIDGET2 FOV and the GRB's position
using a special script. Since WIDGET covers 33\% of the Swift/BAT
FOV., some of events are out of its FOV. If the GRB is in the FOV, the
data sets for the GRB field before and around the GRB trigger time are
manually transferred from the Kiso Observatory to RIKEN and Saitama
University via the Internet.  As summarized in table \ref{obslog1},
all three WIDGET systems (WIDGET 1.0, 1.5 and 2.0) have hunted 11 GRB
fields before their trigger event, except in unfavorable weather
conditions. The time coverage of pre-trigger monitoring is fully
dependent on the satellite observing strategy.  In case of {\it
  HETE-2} (anit-solar pointing), WIDGET was able to obtain long
monitoring data such as for GRB050408 \citep{050408}, GRB051028
\citep{051028} and GRB060121A. In the {\it Swift} era, the time
coverage is shorter (typically several ten minutes) due to {\it
  Swift}'s short pointings.

\section{Calibration and System Performance}

System evaluation is performed using daily data sets. The main
concerns are astrometric and photometric calibration to enable prompt
optical monitoring of GRBs. Since joint observation and analysis with
larger aperture follow-up telescopes (e.g., the Kiso Schmidt
telescope; the Lulin One meter telescopes) are also one of the key
issues for GRB sciences, the magnitude conversion to standard
photometric filters is critical.  Another crucial application is as a
new transient surveyor which can search for highly variable optical
transients (e.g. OT J004240.69+405142.0 \citep{ot}) utilizing entire
WIDGET2 images.  For effective analysis of the huge amounts of data,
the process of astrometric pipeline reduction is essential.

A standard preprocessing routine based on IRAF is employed, including
bias/dark subtraction, and flat-fielding corrections with appropriate
calibration data.

\subsection{Astrometric Calibration}

Astrometric calibration is made using daily observational data.  The
main goal is to identify GRB positions with less than 1 pixel position
accuracy.  The typical position accuracies are $2\sim3"$ for X-ray
({\it Swift}/XRT) and less than $1"$ for optical data,
respectively. This position accuracy requires astrometric accuracy of
less than 1 pixel size of WIDGET for identification of the GRB
position. This effort is also critical to reduce contamination from
other sources.

The geometric distortion effect from the lens optics is well described
by a polynomial function \citep{guan,hata}. According to the generic
features, the effect can be ignored at the center of the images, but
becomes more significant with the distance from the optical axis.  In
order to check the effect on the astrometry of the WIDGET2 images, we
performed astrometry using the central one-ninth part of the image
expected to be affected by the non-geometric. For the astrometric
standard, we employ the Tycho2.0 catalog, which is particularly suited
to use for astrometric use in the point of sky coverage with bright
enough stars (V$_{T}<11.5$) for WIDGET and for position accuracy ($\pm
0.6$ mas).  We selected 21 non-saturated stars brighter than V$_{T}<9$
mag in the field-of-view then identify their coordinates from the
Tycho2.0 catalog. The X-Y positions of each star in the frame are
measured using the IRAF {\tt center} command. Using ccmap of IRAF
task, we calculate the translation form with a tangent sky geometry
projection.  From this fitting, we obtain the pixel scale and field of
view are $55"$ and $10.^\circ7 \times 10.^\circ7$, respectively.  The
spatial resolution and the field-of-view are consistent with the
expected values from the optics and CCD combination, respectively.  In
addition, the position accuracy satisfies our scientific requirement.

\begin{figure}
  \begin{center}
    \FigureFile(80mm,80mm){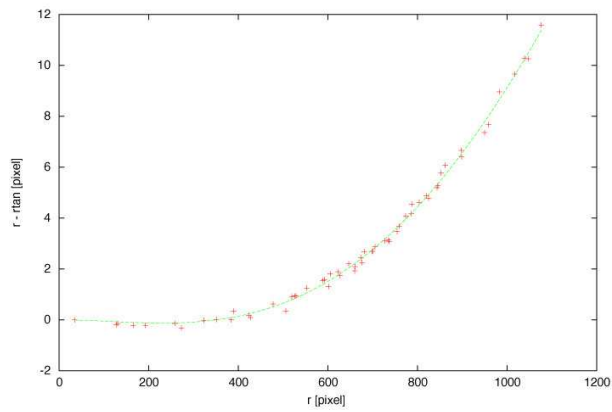}
  \end{center}
  \caption{Optical geometric distortion of the WIDGET lens. The dashed line shows the geometrical solution (see text).}\label{dist}
\end{figure}

We evaluated the differences between actual positions of catalogued
stars on the images and expected coordinates calculated using the
translation function. Figure \ref{dist} shows the differences along
the interval from the optical axis. A geometrical model of field
distortion using a 4th-order polynomial function is employed to
describe this,
\begin{equation}
r - r_{tan} = ar + br^2 + cr^3 + dr^4,
\end{equation}
where $r$ and $r_{tan}$ are distances from the optical axis in units of
pixels for catalog star positions on the CCD images and expected
position as obtained by the translation function. As shown in figure
\ref{dist}, the difference can be well fitted according to equation 1.
This result is consistent with those employing other strategies for a
simple lens system (e.g. \citet{guan,hata}).

We then correct the distortion using the IRAF {\tt GEOMAP} and {\tt
  GEOTRAN} tasks and perform astrometry over the full image area in
the same manner as described above. The astrometric accuracy is
much improved with 45'' rms deviation, similar to the value
for the central region.

\subsection{Photometric Calibration}

On the CCD chip the U10 as peak QE curve is at 720 nm which is similar
to the band-bass of Johsonn-Cosinus Rc-band and SDSS $r'$/$i'$
filters.  It is expected that the best strategy for photometric
calibration is to make a comparison with the Rc-band or $r'$/$i'$-band
brightness.  This is also same assumption with the unfiltered
telescope system \citet{henden}.  In addition, this conversion to $Rc$
or $r'$ bands allows us to connect with further afterglow
observations.  As a first step, we check the variance from the
Hipparcos B$_{\rm T}$ and V$_{\rm T}$ band catalog magnitudes recorded
in the Tycho-2.0 catalog using daily data sets. Since Tycho-2.0 covers
the whole sky and contains bright stars (m$<9$ mag), it is easy to
find many stars on the WIDGET images. Even using V$_{\rm T}$ band, the
scatter is about $0.2\sim0.3$ in magnitude.  This is because
combination effect of the color dependence of the CCD response and
various spectrum of photometric comparison stars.  Since the spectrum
of the prompt optical emission is totally unknown, this calibration
method could have a large systematic error.

For better calibrations, we employ two photometric catalogs; (1)
Landolt standard stars; and (2) Guide star photometric catalog II.
The tentative limiting magnitude of the current system derived from
the Tycho2.0 V$_{\rm T}$-band is deep enough to make a comparison with
these two catalogues, including the redder bands.  These two
photometric catalogs are appropriate for WIDGET data calibration.  The
Landolt catalogue has been used for current optical observations even
for 8-m class telescopes.  The GSPC-II catalog was created to provide
photometric calibrators for the HST Guide Star Catalog. These stars
are randomly distributed throughout the whole sky with huge volume
being 9 mag in the $Rc$-band.  In order to perform additional
photometric calibration, we obtain images centered on the Landolt
standard star field (SA95).  This is in addition to normal daily
observations with 5, 60, 120 and 180 s exposures made under remote
operation from the RIKEN site on 4th November 2007.  The WIDGET field
of view is wide enough to image the entire SA95 field together with
the bright spectroscopic standard star HR1544. Five Landolt's standard
stars (95-15, 95-43, 95-74, 95-96 and 95-149) can be detected in the
images with 5 s exposures. The faintest one is 95-74 ($R=10.93$).  The
color range of $(B-V)$ is from 0.1 to 1.89, which covers the typical
colors of optical afterglow.  These stars are located at similar
airmass 1.12 for HR1544 and 1.27 for the SA95 field, respectively. We
also selected GSPC stars at the same airmass (1.15) with HR
1544. Therefore, the expected differences of atmospheric extinction
for these stars are 0.01 mag level which is not significant for this
calibration.

\begin{figure}
  \begin{center}
    \FigureFile(80mm,80mm){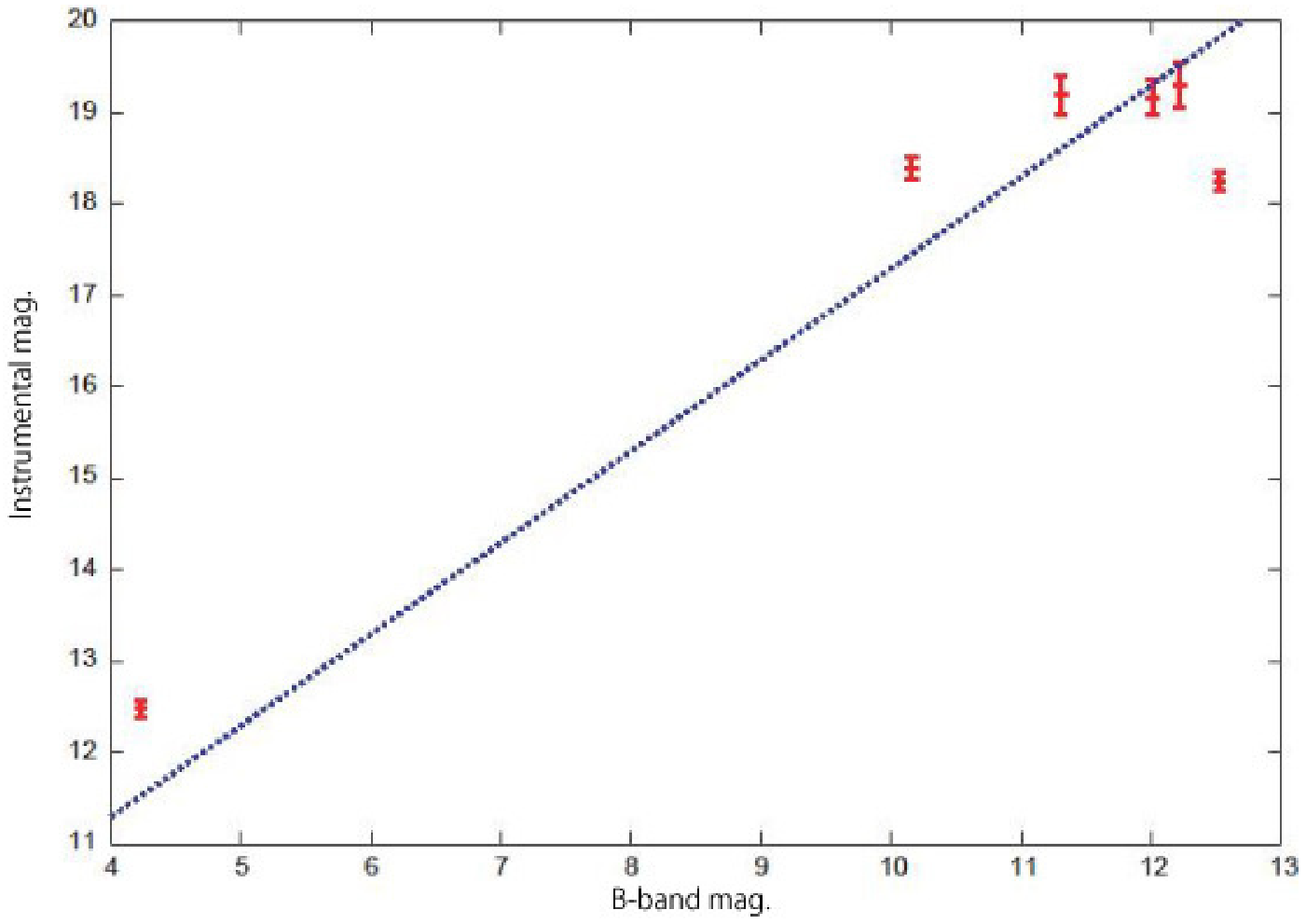}
    \FigureFile(80mm,80mm){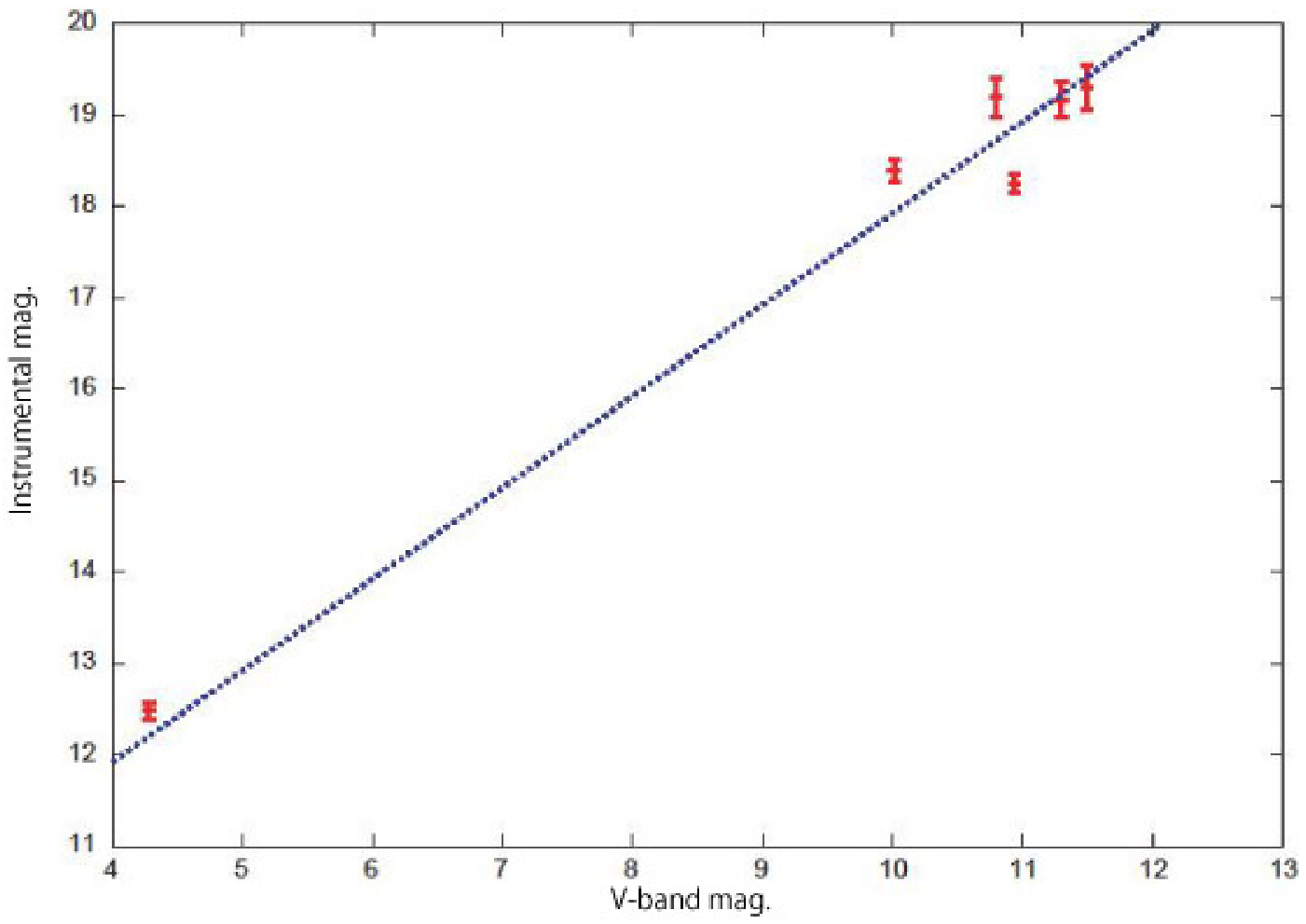}
    \FigureFile(80mm,80mm){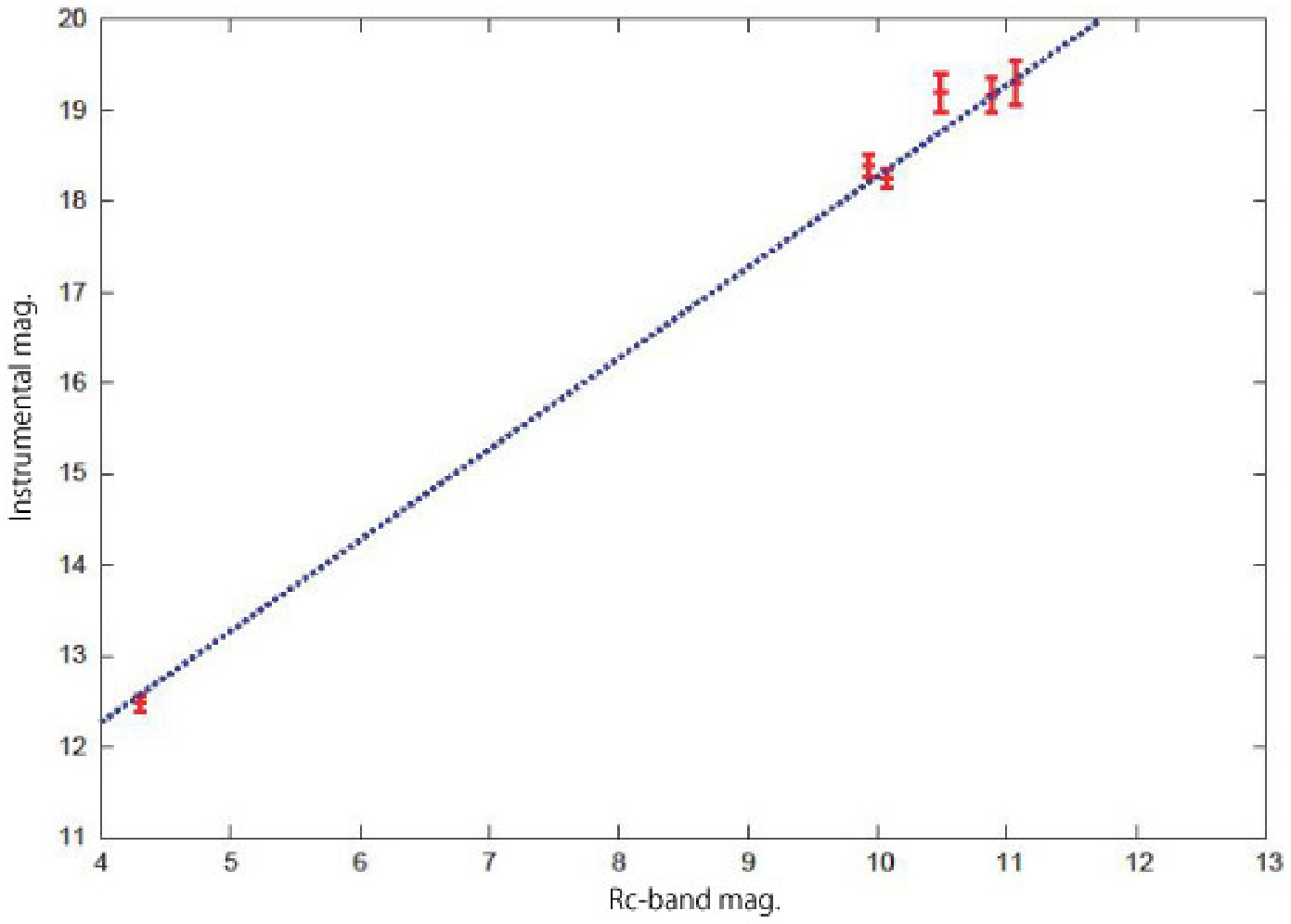}
    \FigureFile(80mm,80mm){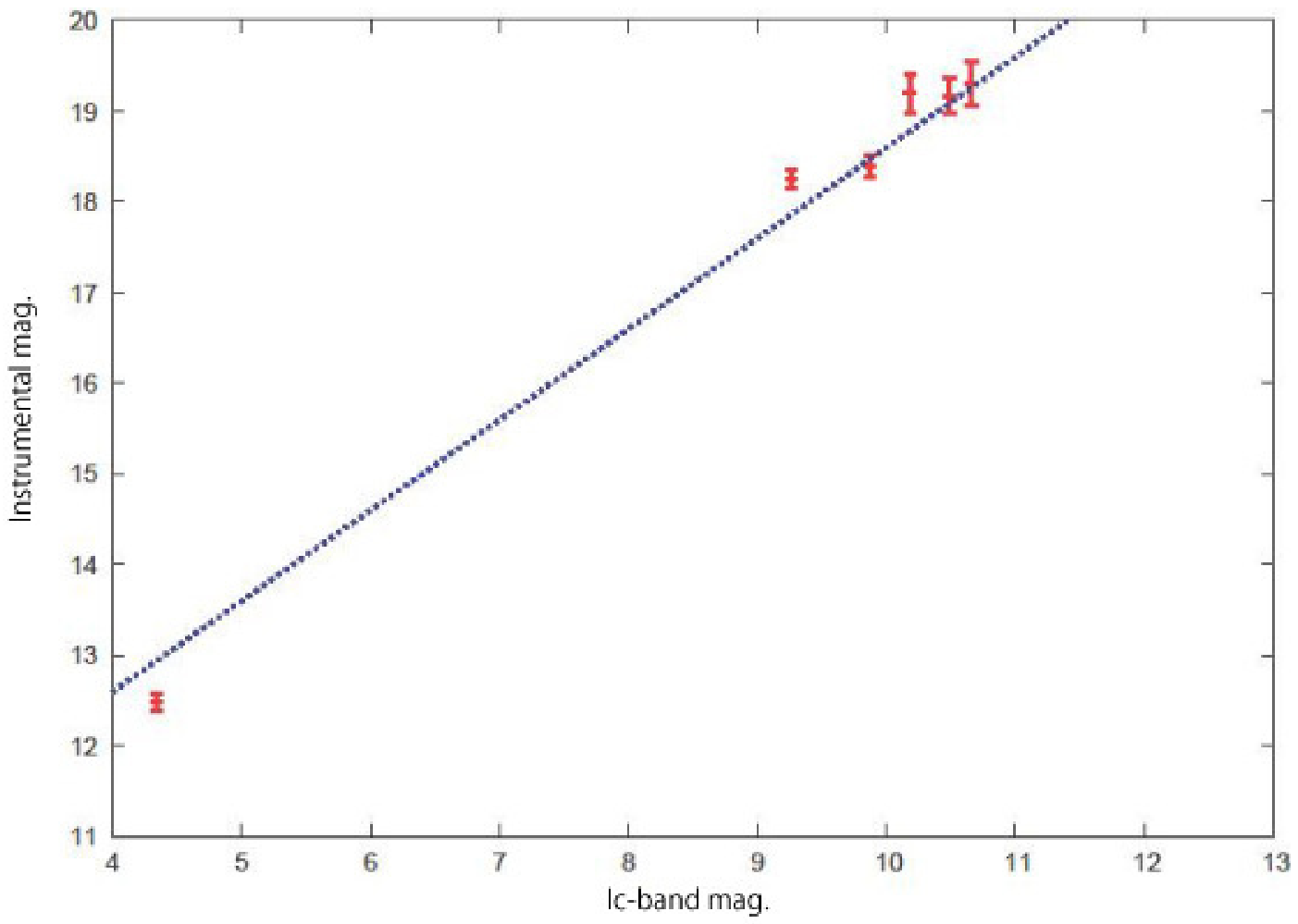}
  \end{center}
  \caption{The photometric comparison of the WIDGET instrumental magnitude with B, V, Rc, Ic magnitudes of the Landolt stars and the spectroscopic standard star HR1544.}\label{landolt}
\end{figure}

\begin{figure}
  \begin{center}
    \FigureFile(80mm,80mm){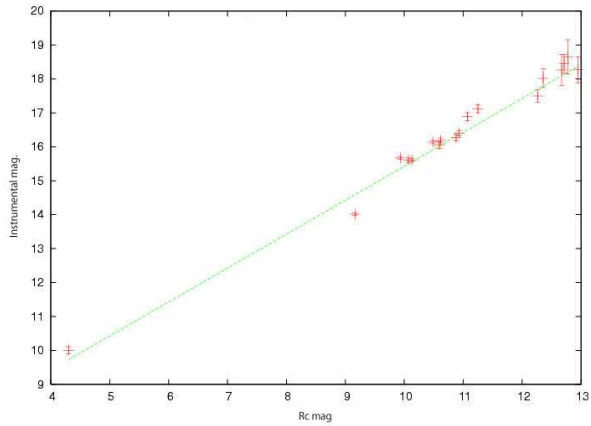}
  \end{center}
  \caption{The R-band comparison of the WIDGET instrumental magnitudes against with the Landolt SA95 field, HR1544 and GSPC catalog.}\label{gspc}
\end{figure}

Based on the standard aperture photometry, we measure the instrumental
magnitude of WIDGET. Figure \ref{landolt} shows a comparison of the B,
V, Rc and Ic magnitudes for the Landolt catalog. The fits made using a
simple liner function yield an rms of 0.50 (B), 0.20 (V), 0.07 (Rc)
and 0.16 (Ic), respectively.  Hence, the WIDGET magnitude is well
matched to standard $Rc$ magnitude.  We also add GSPCII stars to
improve the statistics and to cover bright end of the magnitude
distribution.  As can be seen in figure \ref{gspc} the fitting remains
consistent with the simple conversion to the $Rc$-band magnitude
within the 0.1 mag scatter.  The typical limiting magnitudes in the
$Rc$-band for each exposure are 11.3 (5 s), 12,9 (60 s), 12.8 (120 s)
and 12.9 (180 s), respectively.

\begin{figure}
  \begin{center}
    \FigureFile(80mm,80mm){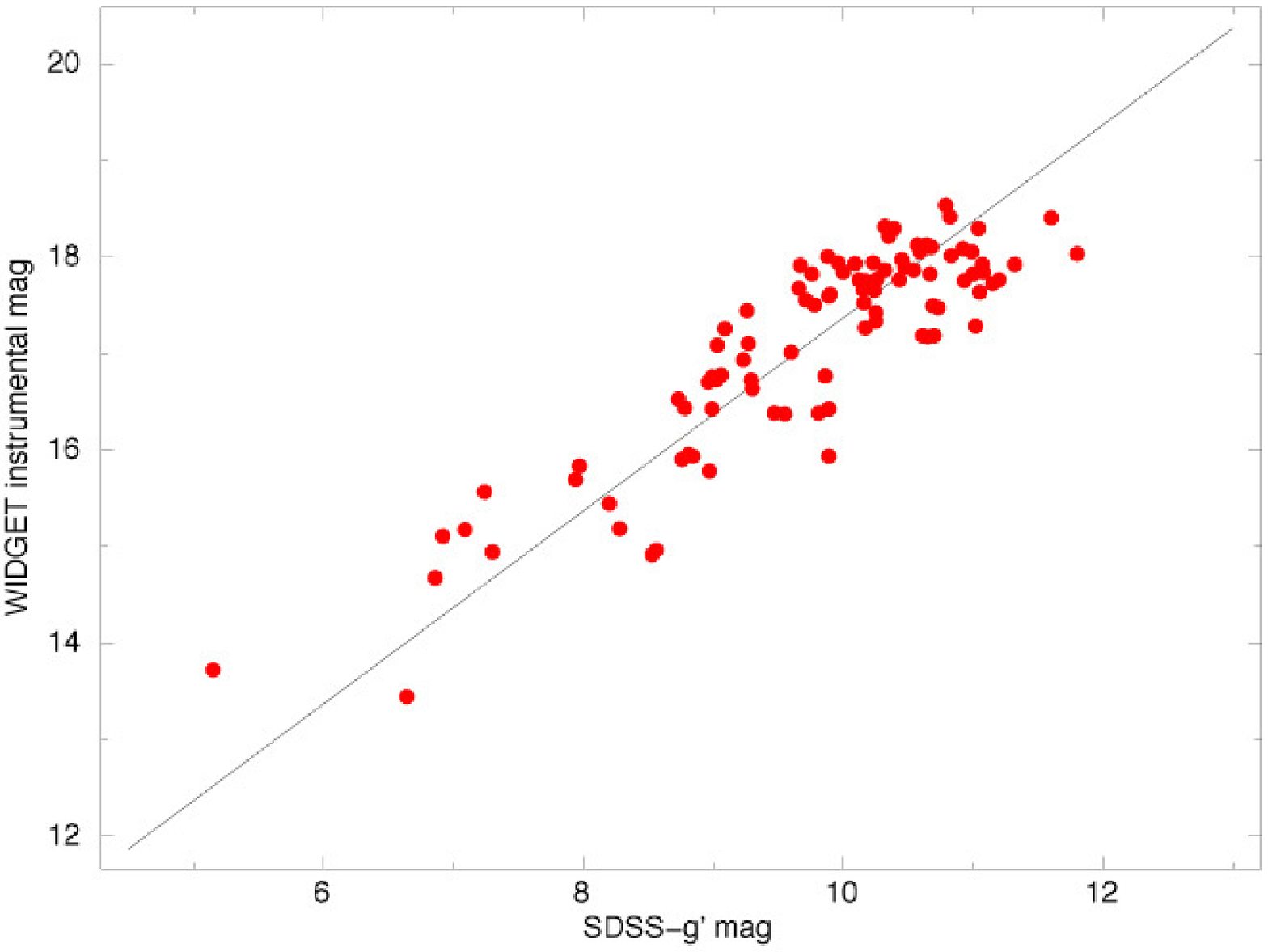}
    \FigureFile(80mm,80mm){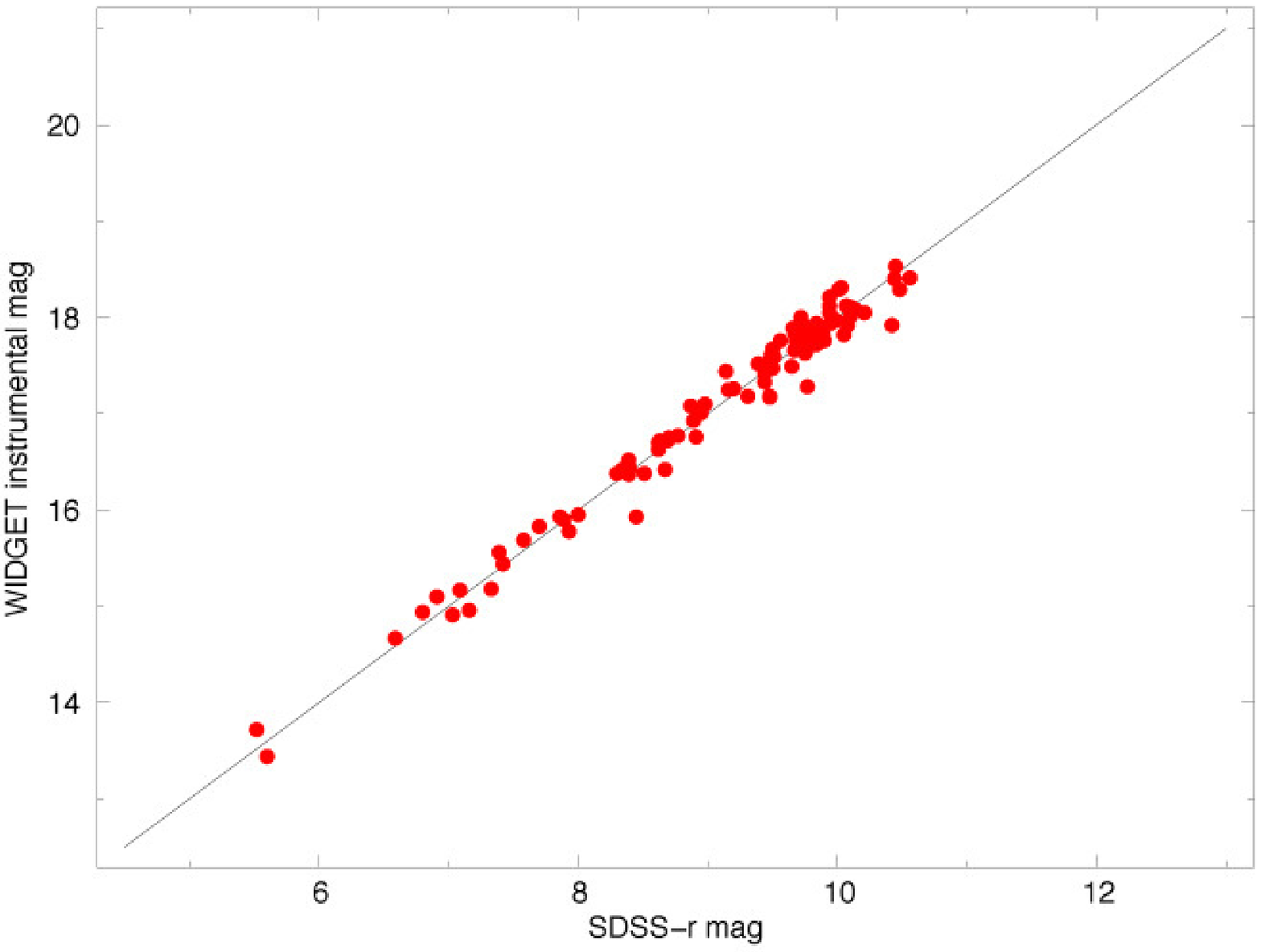}
    \FigureFile(80mm,80mm){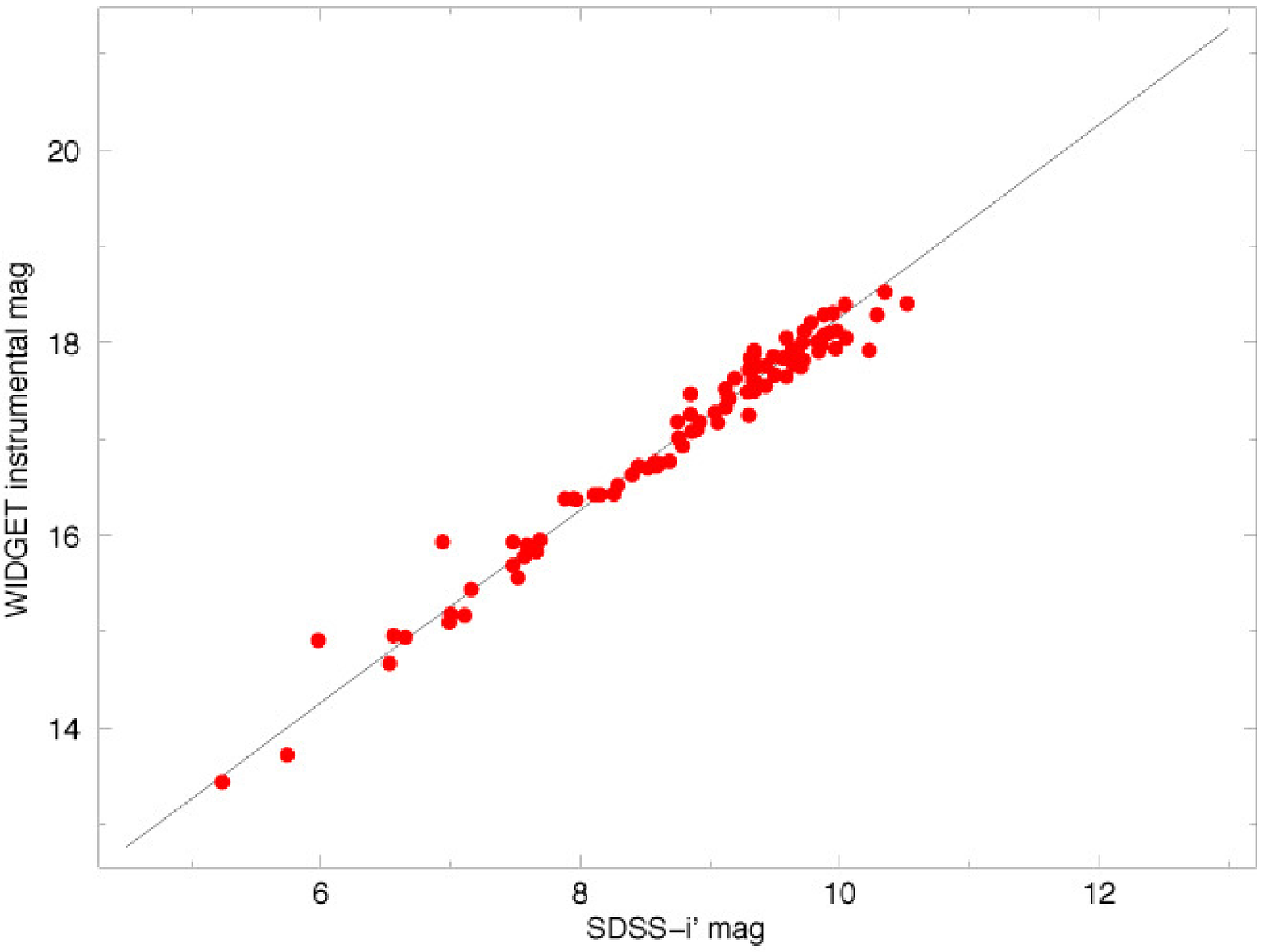}
    \FigureFile(80mm,80mm){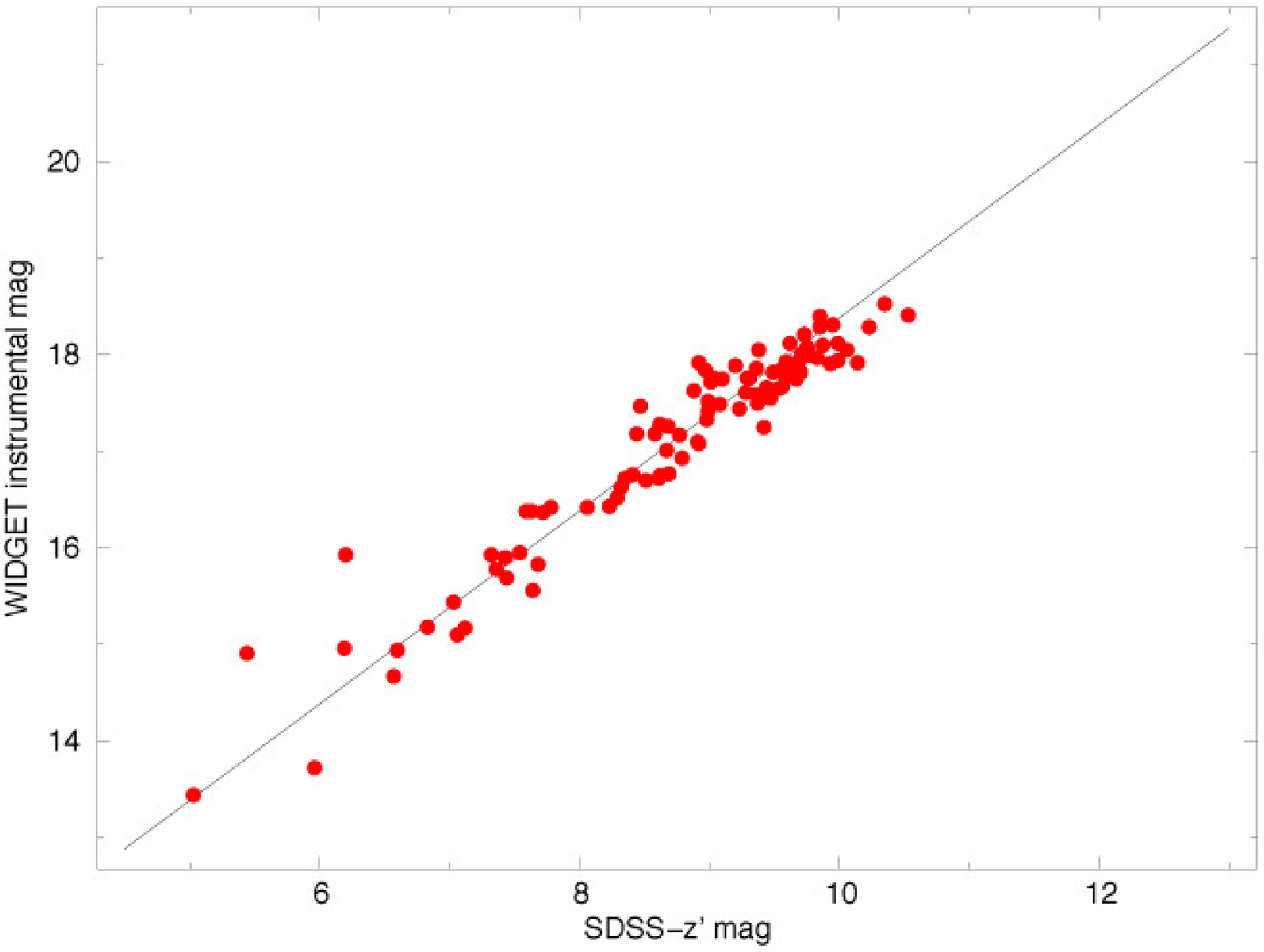}
  \end{center}
  \caption{The photometric calibration against with the Tycho-2.0
    catalog translated to the SDSS g',r',i' and z' band
    system.}\label{tycho-sdss}
\end{figure}

\begin{figure}
  \begin{center}
    \FigureFile(80mm,80mm){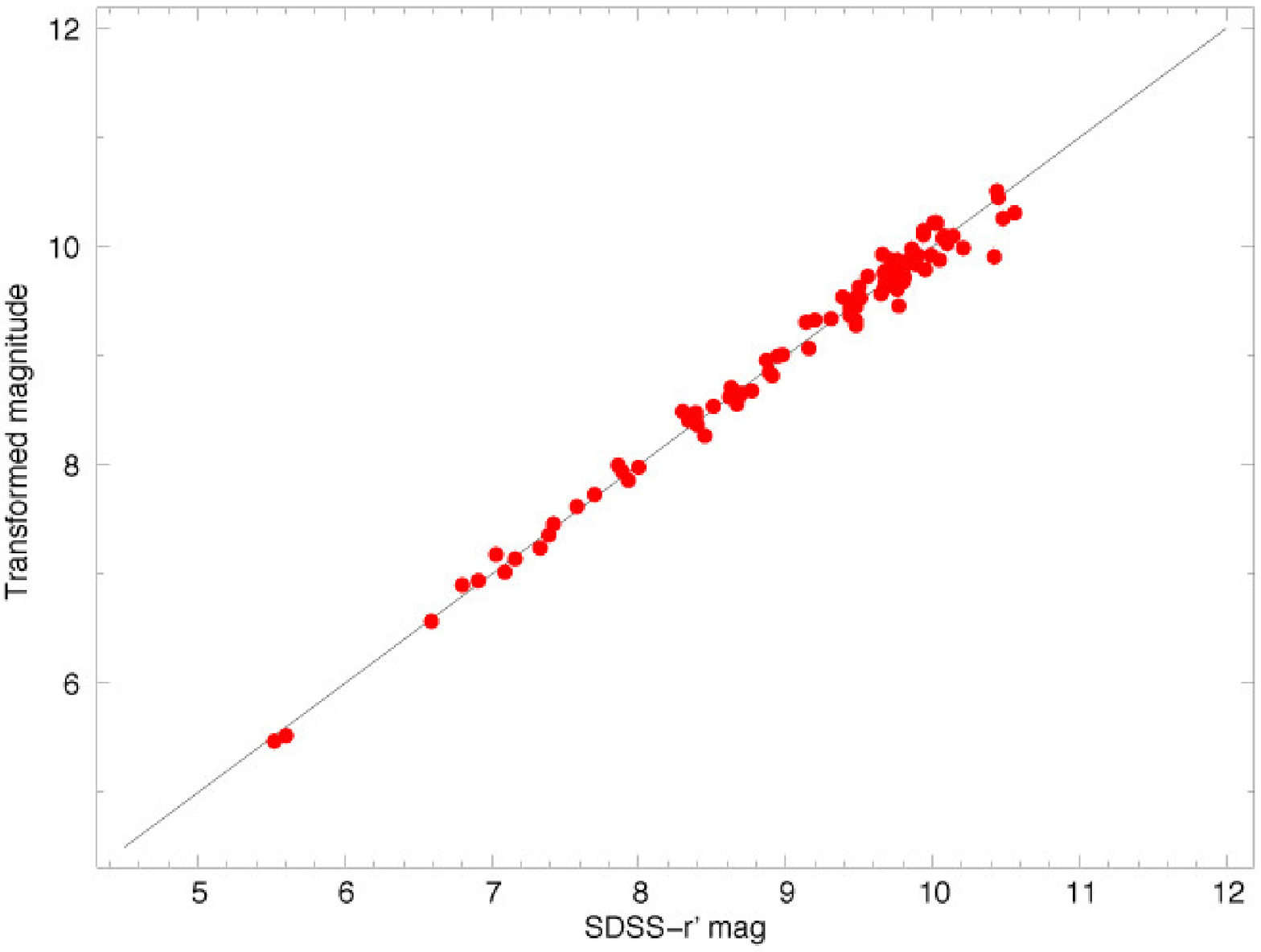}
  \end{center}
  \caption{The photometric calibration against with the Tycho-2.0
    catalog translated to the SDSS r' band system.}\label{color}
\end{figure}

We also make a comparison with the SDSS system using the modified
Tycho-2.0 catalog, as described by \citet{tycho-sdss}.  This is needed
because the number density of bright ($<12$ mag) Landolt and GSPCII
stars is insufficient to perform photometric calibration for numerous
GRB fields.  Typically, there are no Landolt and only a few GSPCII
stars located near the GRB fields.  \citep{tycho-sdss} constructed an
all-sky catalog of griz magnitudes of bright stars ($V < 12$) based on
the Tycho-2.0 catalog, which makes it easy to find a larger number of
bright stars around GRB positions on WIDGET images.  This helps to
minimize the photometric systematic error due to the spatial
differences between references and GRB positions.  Figure
\ref{tycho-sdss} shows the results obtained in the same manner with
Landolt and GSPSII calibrations.  The rms scatter of transformation
between catalog magnitude and instrumental magnitude is 0.59 mag at
$g'$, 0.17 mag at $r'$, 0.19 mag at $i'$ and 0.32 mag at $z'$,
respectively.  It can be seen that the instrumental magnitudes are
sufficiently well reproduced for $r'$ and $i'$ bands.  We also
formulate an equation for transformation from the instrumental
magnitude to the $r$'-band system utilizing color information:
$r' = m_{inst} - 8.15 + 0.17\times(g'-i')$.
Here, $m_{inst}$ indicates the instrumental magnitudes derived with
the pseudo zero-magnitude as 25.00 mag.  Figure \ref{color} shows a
comparison of the instrumental magnitudes transformed
onto the standard system. The scatter becomes 0.12 mag which is 0.05
mag better than the initial result.

\section{Simultaneous optical and $\gamma$-ray observations of GRBs}

We collected at total of ten simultaneous GRB observations, including
one short GRB and one pre-trigger observations, for GRB060121A.  Three
of those bursts were localized by {\it HETE-2}, and the remaining
eight are {\it Swift} events.  Table \ref{obslog1} shows a summary of
the fluence of the $\gamma$-ray emission, duration, redshift and
afterglow brightness. Unfortunately, there is no GRB with an optical
afterglow as bright as GRB030329 or GRB080319B.  These fluences of
those bursts are also an order of magnitude below that of GRB080319B.
Since WIDGET is the only instrument covering the Asian region, there
are no coincident observations from similar instruments such as with
``Pi of the Sky (Las Capanas)'', TORTORA (La Silla) or RAPTOR (Fenton
Hill, US) which cover the Western hemisphere. Therefore, we missed the
extremely bright optical emission associate with GRB080319B because it
occurred in the afternoon in Japan.  Compared with the latter three
instruments, WIDGET has the widest field of view, as summarized in
table \ref{comp}.  This means that the number of simultaneous
observations by WIDGET is about twice that of similar instruments such
as ``Pi of the Sky'' \citep{pi}.  There are also seven {\it HETE-2}
events that occured under cloudy conditions. When we include cloudy
events, the simultaneous observational ratio for the {\it HETE-2}
events is 6.7 GRBs/yr, which is consistent with the expected ratio and
four times higher than that of {\it Swift}.  One of the main reasons
for the significant differences between the {\it HETE-2} and {\it
  Swift} ratios is the pointing strategy.  {\it HETE-2} monitored the
anti-solar direction, while the distribution of {\it Swift} pointings
is mostly uniform. Anti-solar monitoring is easier to achieve for
ground-based telescopes.  The next generation GRB satellite SVOM will
explore the anti-Sun direction and is expected to localize about 70
GRB per year (Schanne et al. 2010). WIDGET will be able to make
3$\sim$4 observations per year by monitoring the SVOM FOV.

All 11 GRB positions in table 3 were covered by WIDGET before the
onset of the $\gamma$-ray emissions.  In order to search for optical
precursors, we checked all available data sets for the presence of
emission at the alerted position evaluating CCD ADU counts, sky
background, and statistical error for time series. The GRB region is
selected using a 2 pixel aperture centered on the afterglow position.
In the GRB050408 case, there was a significant emissions besides
cosmic-ray events 4.7 hrs before the trigger. However, the object was
moving just passing through the GRB position. Except for this moving
object, there was no significant optical emission.  Besides GRB050408,
there are no other significant detections at the other ten GRB
positions.  Note that the X-ray afterglows of four events (GRB051227,
GRB060211A, GRB060413 and GRB071021) show a clear plateau
phase. According to \citet{yamazaki}, there is a possible optical
precursor $10^{3}-10^{4}$ s before the trigger. We did not detect any
significant emission from these events in the time region summarized
in table \ref{obslog1}.

\begin{figure}
  \begin{center}
    \FigureFile(80mm,80mm){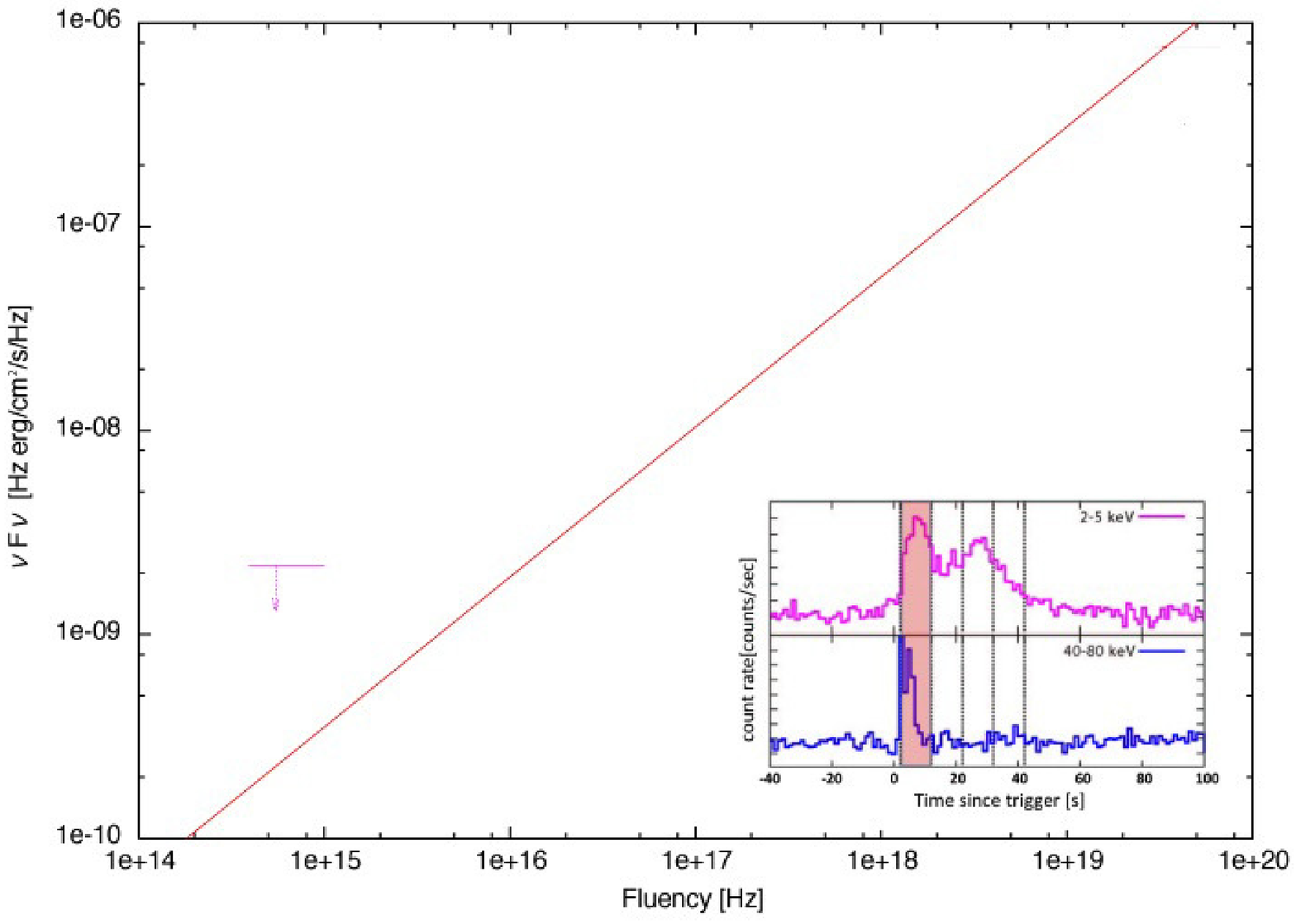}
    \FigureFile(80mm,80mm){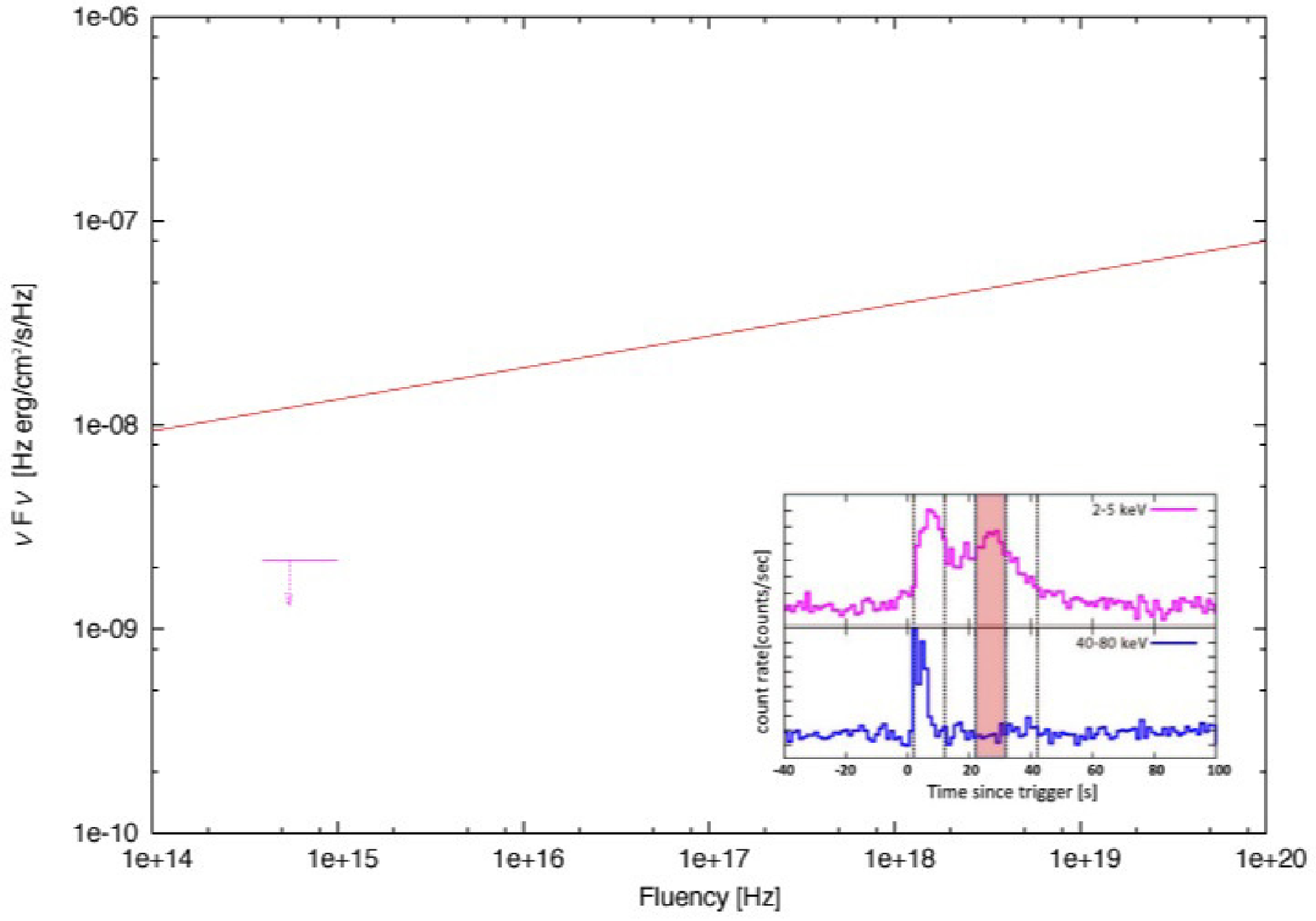}
  \end{center}
  \caption{Time resolved broad band SED of the GRB050408 prompt emission. Sub panels show the $\gamma$-ray and X-ray light curves. The red regions indicate the time region which we derive the SED. The lines show the best fit function of the low energy power-law.}\label{050408}
\end{figure}

\begin{figure}
  \begin{center}
    \FigureFile(80mm,80mm){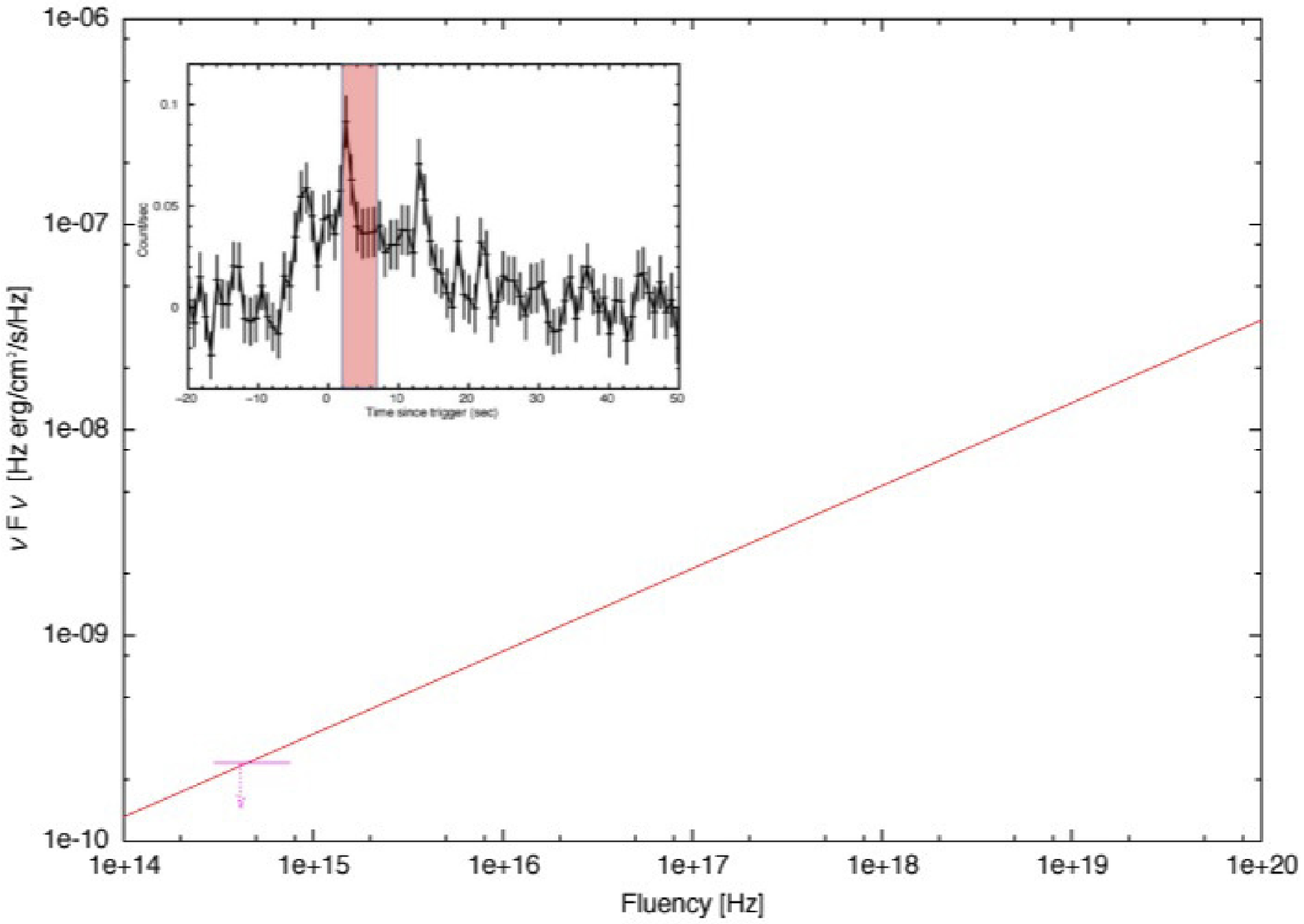}
    \FigureFile(80mm,80mm){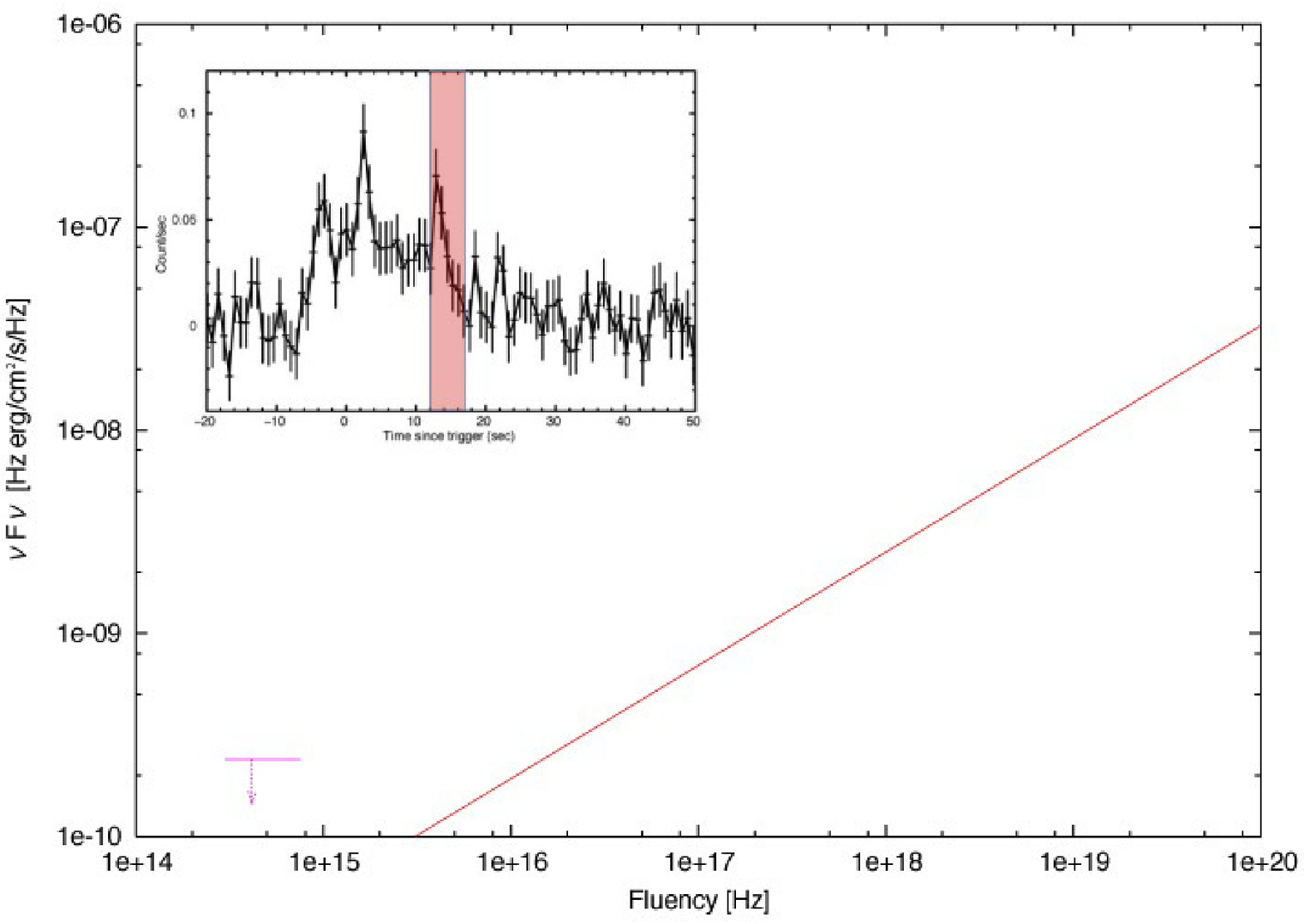}
  \end{center}
  \caption{Time resolved broad band SED for the GRB060323 prompt emission.  Sub panels show the $\gamma$-ray and X-ray light curves. The red regions indicate the time region which we derive the SED. The lines show the best fit function of the low energy power-law.}\label{060323}
\end{figure}

We also constructed a broadband SED that includes $\gamma$-ray and
X-ray data from {\it HETE-2}, {\it Swift} and {\it Suzaku}/WAM. Since
GRB060413 has high galactic extinction ($A_{V}$=6.3), it is excluded
from this joint analysis.  Analysis shows that there is no bright
optical emission similar to that from GRB080319B (V$\sim6$ mag.).
This implies that events like GRB 080319B as the extreme prompt
behavior is understood by synchrotron self-Compton are statistically
rare.  Figures \ref{050408} and \ref{060323} show the time resolved
SEDs for GRB050408 and GRB060323, respectively.  These SEDs show that
the WIDGET limits are below or marginally consistent with
extrapolations from the low-energy power law of the prompt
$\gamma$-ray spectrum.  The absence of the optical emission can be
explained assuming that the prompt $\gamma$-ray and X-ray emissions
are the same synchrotron radiation with the synchrotron
self-absorption between X-ray and optical bands
($\nu_{m}<\nu_{opt}<\nu_{a}$ or $\nu_{opt}<\nu_{a}<\nu_{m}$ case).
Here, $\nu_{m}$, $\nu_{opt}$ and $\nu_{a}$ are the frequencies of the
observed characteristic emission, optical band and self-absorption,
respectively.  Another possibility is the $\nu_{a}<\nu_{opt}<\nu_{m}$
case (for details see \citet{shen}).

\section{Conclusions}

We summarize the system development, operation, as well as GRB observations.

\begin{itemize}

\item Fully robotic instruments for observation of prompt optical
  emission associated with GRB were constructed. The first system
  monitored the {\it HETE-2} field of view between 2004 and 2007. The
  current system is making simultaneous monitoring for {\it Swift}/BAT
  with 5-second exposures.

\item The geometric distortion of the optical system is evaluated, and
  astrometric calibration against with the Tycho-2.0 catalog is
  performed. The distortion is well described with a 4th order
  polynomial function. After the distortion correction, the
  astrometric accuracy is less than 1 pixel level.

\item Photometric calibration against with the Landolt photometric
  catalog and modified Tycho-2.0 catalog is performed. WIDGET's
  unfiltered magnitude is well converted to the SDSS r' system at a
  10\% level.

\item WIDGET made 11 GRB observations preceding the $\gamma$-ray
  trigger.  The efficiency of the synchronized observations with {\it
    HETE-2} is four times better than that of {\it Swift}. The time
  resolved SEDs for two GRBs (GRB050408 and GRB060323) show that the
  WIDGET limits are below or marginally consistent with the
  extrapolations of the $\gamma$-ray spectrum. The absence of the
  optical emission can be explained assuming that the prompt
  $\gamma-$-ray and X-ray emissions are the same synchrotron radiation
  with the self-absorption between X-ray and optical bands.

\end{itemize}

\bigskip

We thank the staff at the Kiso Observatory for the various
arrangements.  The project is supported by the RIKEN director's fund
in FY2003 and FY2004 (PI TT) and a Grant-in-Aid from the Ministry of
Education, Culture, Sports, Science and Technology of Japan (1834005;
PI: MST, 18001842; YU). This work is partly supported by grants NSC
98-2112-M-008-003-MY3 (YU).


\clearpage

\begin{table}
  \caption{WIDGET systems}\label{system}
  \begin{center}
    \begin{tabular}{llllll} \hline
Version & FOV. &  Lens & Pixel size & Site &  Operation period \\ \hline \hline
WIDGET1.0 & $(62^{\circ} \times 62^{\circ}) \times 1$&Cannon EF 24mm f/1.4& $1'.82\times1'.82$ & Akeno & 2004.07--2005.11\\
WIDGET1.5 & $(42^{\circ} \times 42^{\circ}) \times 3$&Cannon EF 35mm f/1.4& $1'.23\times1'.23$ & Akeno & 2005.11--2006.11\\ 
WIDGET2.0 & $(32^{\circ} \times 32^{\circ}) \times 4$&Cannon EF 50mm f/1.2& $0'.94\times0'.94$ & Kiso  & 2007.02--current\\ \hline
    \end{tabular}
  \end{center}
\end{table}

\clearpage

\begin{table}
  \caption{Comparison with other system}\label{comp}
  \begin{center}
    \begin{tabular}{llllll} \hline
Name & Site & FOV & Limit & System & Ref\\
\hline
WIDGET        & Kiso, Japan      & $(32^{\circ}\times32^{\circ})\times 4$ & 12  & CCD & This work\\
Pi of the sky & Las Capanas      & $(20^{\circ}\times20^{\circ})\times 2$ & $\sim12.5$    & CCD & \citet{pi} \\
TORTORA       & La Silla         & $30^{\circ}\times24^{\circ}$           &  $10\sim11$   & TV-CCD & \citet{tortora}\\
RAPTOR-Q      & Fenton Hill, USA & All sky                              & 9.5    & CCD & \citet{raptor} \\
RAPTOR-P      & Fenton Hill, USA & $(8^{\circ}\times8^{\circ})\times 4$   & 15 & CCD & \citet{raptor} \\
\hline
    \end{tabular}
  \end{center}
\end{table}

\clearpage

\begin{table}
  \caption{GRB Observation summary}\label{obslog1}
  \begin{center}
    \begin{tabular}{lllllllll} \hline
      GRB    & Localization & Time coverage & $z$ & T90 [s]& Fluence$^{*}$ & Earliest follow-ups & OT & System \\ \hline \hline
      050408 & {\it HETE-2} & $-6.0$ h-- $+3.4$ m  & 1.24 & 19.9  & 50.5$^{\dagger}$    & ROTSEIIIa (18.5s)   & Lulin $R=20.3$ (55 m) & WIDGET-1\\ 
      051028 & {\it HETE-2} & $-16.0$ m -- $+11.2$ m  & $--$ & 18.8  &  60$^{\dagger}$     & Lulin (2.28h)       & $R=20.8$        & WIDGET-1\\ 
      051227 & {\it Swift}  & $-187$ m -- $+12$ m    & 0.71 & 114.6 &  7.0$^{\ddagger}$     & RIMOTS (20 min)     & VLT $R\sim25$ (10.3h) & WIDGET-1 \\
      060121A& {\it HETE-2} & $-8$ h -- $-3.5$ h  & $--$ &  3.6  &  70.2$^{\dagger}$    & NOT (1.98h)         & $R=22.65$             & WIDGET-1\\
      060211A& {\it Swift}  & $-13.2$ m -- $+5.4$ m    & $--$ & 126.3 &  15.7$^{\ddagger}$   & ROTSEIIIA (147s)    & $R<14.3$              & WIDGET-1.5\\
      060323 & {\it Swift}  & $-12.5$ m -- $+1.4$ m  & $--$ &  25.4 &  6.2$^{\ddagger}$    & Xinglong (540s)     & $R=18.2$              & WIDGET-1.5\\  
      060413 & {\it Swift}  & $-0.4$ m -- $+5.6$ m   & $--$ & 147.7 &  35.6$^{\ddagger}$   & UVOT (124s)         & $V<19.2$              & WIDGET-1.5\\
      070616 & {\it Swift}  & $-0.5$ m -- $+2.7$ m   & $--$ & 402.4 &  192.0$^{\ddagger}$  & MITSUME (23m)       & BTA $R\sim22$ (7.5h)  & WIDGET-2.0\\
      070810B& {\it Swift}  & $-1.2$ m -- $20.1$ m   & $--$ & 0.08  &  0.12$^{\ddagger}$   & Xinglong (300s)     & $R<20.5$              & WIDGET-2.0\\
      071021 & {\it Swift}  & $-13.4$ m -- $+140.8$ m & $--$ & 225   &  13$^{\ddagger}$     & Faulkes (258s)      & $R=17.8$              & WIDGET-2.0\\
      090408 & {\it Swift}  & $-20.0$ m -- $+3.0$ m  & $--$ & $--$  &  $--$   & TAOS (91s)          & $R<15.2$              & WIDGET-2.0\\
\hline
$^*$:$10^{-7} {\rm erg~cm^{-2}}$\\
$^{\dagger}$: 30-400 keV band\\
$^{\ddagger}$: 15-150 keV band\\
    \end{tabular}
  \end{center}
\end{table}


\begin{thebibliography}{}


\bibitem[Akerlof et al.(1999)]{990123} Akerlof, C., et al.\ 
1999, \nat, 398, 400 

\bibitem[Bloom et al.(2009)]{bloom} Bloom, J., et al.\ 2009, \apj, 691, 723

\bibitem[de la Fuente Marcos \etal (2009)]{ot} de la Fuente Marcos, R., et al.\ 2009 New Astronomy 14, 214

\bibitem[de Ugarte Postigo et al.(2007)]{050408} de Ugarte Postigo, A., et al.\ 2007, \aap, 462, L57 

\bibitem[Greco \etal (2009)]{tortora} Greco, G., et al. \ 2009, Mem. S.A.It. Vol 80, 231

\bibitem[Guan \etal(1997)]{guan} Guan, H., Aoki, S., \& Ejiri, K., 1997, Richoh Technical Report No23. 48

\bibitem[Hata \etal (2002)]{hata} Hata, K., Maruyama, M., \& Toriihara, M., Oobayashi gumi technical report 2002, No. 65, 71

\bibitem[Hayashida et al.(1996)]{agasa} Hayashida, N., et 
al.\ 1996, Physical Review Letters, 77, 1000 

\bibitem[Henden (2000)]{henden} Henden, A. A., 2000, Journal AAVSO vol. 29, page 35.

\bibitem[Huang et al.(2005)]{040924} Huang, K.~Y., et al.\ 
2005, \apjl, 628, L93 

\bibitem[Huang et al.(2007)]{050319} Huang, K.~Y., et al.\ 
2007, \apjl, 654, L25 


\bibitem[Kobayashi(2000)]{re1} Kobayashi, S.\ 2000, \apj, 
545, 807 

\bibitem[Kobayashi 
\& Sari(2000)]{re2} Kobayashi, S., \& Sari, R.\ 2000, \apj, 542, 819 

\bibitem[Ofek (2008)]{tycho-sdss} Ofek, E. O, 2008, PASP, 120, 1128

\bibitem[Page et al.(2007)]{061121} Page, K.~L., et al.\ 2007, 
\apj, 663, 1125 

\bibitem[Racusin et al.(2008)]{080319b} Racusin, J.~L., et al.\ 
2008, \nat, 455, 183 

\bibitem[Schanne et al.(2010)]{svom} Schanne, S., et al.\ 
2010, arXiv:1005.5008 

\bibitem[Shen \& Zhang(2009)]{shen} Shen, R.-F., \& Zhang, B.\ 2009, \mnras, 398, 1936 

\bibitem[Sokolowski et al.(2009)]{pi} Sokolowski, M., et 
al.\ 2009, American Institute of Physics Conference Series, 1133, 306 

\bibitem[Tamagawa \etal(2005)]
 {tamagawa}Tamagawa, ~T. \etal\ 2005, NCimC, 28, 771

\bibitem[Urata et al.(2007a)]{sum} Urata, Y., et al.\ 2007, 
\apjl, 668, L95 

\bibitem[Urata et al.(2007b)]{051028} Urata Y., et al.\ 2007, PASJ, 59L, 29

\bibitem[Urata et al.(2007c)]{041006} Urata, Y., et al.\ 2007, 
\apjl, 655, L81 

\bibitem[Urata et al(2005)]{eafon}
  Urata Y. et al\ 2005,  NCimC, 28, 775

\bibitem[Urata et al.(2004)]{030329} Urata, Y., et al.\ 2004, 
\apjl, 601, L17 

\bibitem[Urata et al.(2003)]{020813} Urata, Y., et al.\ 2003, 
\apjl, 595, L21 

\bibitem[Vestrand et al.(2006)]{041219a} Vestrand, W.~T., et 
al.\ 2006, \nat, 442, 172 

\bibitem[Vestrand et al.(2005)]{050820a} Vestrand, W.~T., et 
al.\ 2005, \nat, 435, 178 

\bibitem[Wozniak \etal (2009)]{raptor} Wozniak P.R., et al.\ 2009, ApJ, 691 495

\bibitem[Yamazaki(2009)]{yamazaki} Yamazaki, R.\ 2009, \apjl, 
690, L118 

\bibitem[Yost et al.(2007)]{051111} Yost, S.~A., et al.\ 2007, 
\apj, 657, 925 

\end{thebibliography}
\end{document}